\documentstyle[prb,aps,epsf,floats]{revtex}
\newcommand{\beq}{\begin{equation}}
\newcommand{\eeq}{\end{equation}}
\newcommand{\s}{{\sigma}}
\newcommand{\psib}{{\bar{\psi}}}
\newcommand{\rb}{{\bar{\rho}}}
\newcommand{\phib}{{\bar{\phi}}}
\newcommand{\dt}{{\Delta}}
\newcommand{\dva}{{\frac{\vp\times\va}{2\pi}-\rb}}
\newcommand{\dvab}{{\frac{\vp\times(\va-\vb)}{4p\pi}}}
\newcommand{\w}{{\omega}}
\newcommand{\zh}{{\hat{z}}}
\newcommand{\qh}{{\hat{q}}}
\newcommand{\vA}{{\vec{A}}}
\newcommand{\va}{{\vec{a}}}
\newcommand{\vrr}{{\vec{r}}}
\newcommand{\vj}{{\vec{j}}}
\newcommand{\vE}{{\vec{E}}}
\newcommand{\vB}{{\vec{B}}}
\def\bA{{\mathbf A}}
\def\bm{{\mathbf m}}
\def\bsig{{\mathbf \sigma}}
\def\bB{{\mathbf B}}
\def\bp{{\mathbf p}}
\def\bI{{\mathbf I}}
\def\bn{{\mathbf n}}
\def\bM{{\mathbf M}}
\def\bq{{\mathbf q}}
\def\br{{\mathbf r}}
\def\bs{{\mathbf s}}
\def\bS{{\mathbf S}}
\def\bQ{{\mathbf Q}}
\def\bs{{\mathbf s}}
\def\bB{{\mathbf B}}
\def\bl{{\mathbf l}}
\def\bPi{{\mathbf \Pi}}
\def\bJ{{\mathbf J}}
\def\bR{{\mathbf R}}
\def\bz{{\mathbf z}}
\def\ba{{\mathbf a}}
\def\bk{{\mathbf k}}
\def\bP{{\mathbf P}}
\def\bg{{\mathbf g}}
\def\bX{{\mathbf X}}
\newcommand{\etab}{\mbox{\boldmath $\eta $}}
\newcommand{\sigmab}{\mbox{\boldmath $\sigma $}}
\newcommand{\vx}{{\vec{x}}}
\newcommand{\vq}{{\vec{q}}}
\newcommand{\vQ}{{\vec{Q}}}
\newcommand{\vd}{{\vec{d}}}
\newcommand{\vb}{{\vec{b}}}
\newcommand{\vp}{{\vec{\partial}}}
\newcommand{\p}{{\partial}}
\newcommand{\gr}{{\nabla}}
\newcommand{\ra}{{\rightarrow}}
\def\lam{\lambda}
\def\dd{d^{\dagger}}
\def\sb{{\bar{s}}}
\def\half{{1\over2}}
\def\third{{1\over3}}
\def\twof{{2\over5}}
\def\threes{{3\over7}}
\def\rhob{{\bar \rho}}
\def\ua{\uparrow}
\def\da{\downarrow}
\def\eqa{\begin{eqnarray}}
\def\eea{\end{eqnarray}}
\begin{document}
\draft \flushbottom \twocolumn[
\hsize\textwidth\columnwidth\hsize\csname
@twocolumnfalse\endcsname
\title{ Quantum Dots with Disorder and Interactions: A Solvable Large-$g$ Limit}
\author{    Ganpathy Murthy$^1$ and R. Shankar$^2$}
\address{$^1$Physics Department, University of Kentucky, Lexington KY 40506-0055\\
$^2$ Department of Physics, Yale University, New Haven CT 06520}
\date{\today}
\maketitle
\begin{abstract}
We show that problem of interacting electrons in a quantum dot
with chaotic boundary conditions is solvable in the $g\to \infty $
limit, where $g$ is the dimensionless conductance of the dot. The
critical point of the $g=\infty$ theory (whose location and
exponent are known exactly) that separates strong and
weak-coupling phases also controls a wider fan-shaped region in
the coupling-$1/g$ plane, just as a quantum critical point
controls the fan in at $T>0$. The weak-coupling phase is governed
by the Universal Hamiltonian and the strong-coupling phase is a
disordered version of the Pomeranchuk transition in a clean Fermi
liquid.  Predictions are made in the various regimes for the
Coulomb Blockade peak spacing distributions and Fock-space
delocalization (reflected in the quasiparticle width and ground
state wavefunction).
\end{abstract}
\vskip 1cm \pacs{73.50.Jt}]

The union of interactions and disorder in electronic systems poses a
nasty problem: Techniques that work when only one or the other is
present fail when they coexist. Here we discuss a class of
problems involving quantum dots (QD's) where the evil twins can be tamed,
thanks to a small parameter $1/g$, $g$ being the dimensionless
conductance of an open QD. The problem is interesting for
experiments in the Coulomb Blockade (CB)
regime\cite{sivan-expt,marcus-expt,kastner-expt,others-expt}.

We consider $d=2$ dots characterized by $E_F$, the Fermi energy,
$E_T$, the Thouless energy (which is the amount by which the
uncertainty principle broadens the electronic energy levels in the
time it takes to traverse the dot), and $\Delta$, the average single
particle level spacing, with $E_F>>E_T>>\Delta$.  $E_T$ also measures
the band around the Fermi energy wherein the energy levels and wave
functions ($g$ in number) may be described statistically\cite{alt1} by
Random Matrix Theory (RMT)\cite{RMT,reviews}.  The randomness here is
due to the chaotic boundary conditions, with the mean free path $l$
equal to the sample size $L$, and $E_T\simeq E_{\tau}\simeq \hbar
v_F/L$, where $E_\tau$ is the level width due to scattering.

Our hamiltonian is:
\begin{eqnarray} H&=&\sum_{\alpha} \psi^{\dag}_{\alpha}\psi_{\alpha}
\varepsilon_{\alpha}+{1 \over 2}\sum_{\alpha \beta \gamma
\delta}V_{\alpha \beta \gamma
\delta}\psi^{\dag}_{\alpha}\psi^{\dag}_{\beta}\psi_{\gamma}\psi_{\delta}\label{hran}
\end{eqnarray}
where $\varepsilon_{\alpha}$ are single-particle levels that obey RMT
statistics, have a mean spacing $\Delta$ and range from $-g\Delta /2$
to $g\Delta /2$. In the following we will supress spin for simplicity,
pointing out how its restoration modifies various results.

Choices for $V_{\alpha \beta \gamma \delta}$ range in previous work
from all of them being independent gaussian variables\cite{2brim} to
the Universal Hamiltonian\cite{brouwer,kurland,univ-ham}, wherein
$V_{\alpha \beta \beta
\alpha}=u_0\ \Delta$, (all others zero). The sole parameter $u_0$
clearly couples to $Q^2$, $Q$ being the total charge. For the spinful
case a term coupling to total ${\vec S}^2$ is also included.

We employ the choice made by Murthy and Mathur\cite{mm} (MM) who
appeal to the Renormalization Group (RG) approach developed by one
of us\cite{rg-shankar}, where one integrates out modes that lie
outside a narrow band centered on the Fermi energy to expose the
low-energy physics.  In a clean system the RG leads to Landau's
Fermi Liquid (FL) parameters\cite{agd} as fixed-point
couplings\cite{rg-shankar}.  With disorder one expects this RG to
work till we come down to $E_{\tau}\simeq E_T$, still leading to
FL couplings since $E_T<<E_F$. At this point disorder is included
exactly by switching from the $\bk$ basis to the disorder basis:
\begin{eqnarray*} \lefteqn{V_{\alpha \beta \gamma \delta}= {\Delta
\over 4}}\\ & & \sum_{\bk \bk'}u (\theta -\theta' ) \left[
\phi^{*}_{\alpha}(\bk ) \phi^{*}_{\beta}(\bk' )
-\phi^{*}_{\alpha}(\bk' ) \phi^{*}_{\beta}(\bk )\right] * (\alpha
\beta \to \delta \gamma ) \end{eqnarray*} where $\theta$ and $\theta'$
are the angles of the momentum vectors $\bk$ and $\bk'$ and $V_{\alpha
\beta
\gamma \delta}$ is just the transcription of the Fermi liquid
interaction $ V_{FL} ={E_T \over 2g} \sum_{\bk \bk'} u (\theta
  -\theta' ) n_{\bk} n_{\bk'}$.  We will Fourier resolve the
  interaction as $u(\theta -\theta' )=u_0 + \sum_{m=1}^{\infty}u_m
  \cos \left[ m (\theta -\theta' )\right] $. The constant $u_0$
  controls $V$'s with nonzero average $
\langle V_{\alpha \beta \beta \alpha }\rangle = {\Delta \over
g^2}\sum_{\bk \bk'}u(\theta , \theta' )=u_0 \Delta$. All couplings
have nonzero fluctuations given by \beq \langle V^{2}_{\alpha
\beta \gamma \delta }\rangle - \langle V_{\alpha \beta \gamma
\delta }\rangle^{2}={\Delta^2 \over
4g^2}\sum_{m=1}u_{m}^{2}.\label{disp} \eeq Couplings with zero
average are discarded in the Universal
Hamiltonian\cite{brouwer,kurland,univ-ham}. Can these couplings be
relevant in the RG sense and dominate low-energy physics, despite
their small size in the original Hamiltonian?  To answer this,
MM\cite{mm} resorted to further fermionic RG\cite{rg-us} {\em
within} $E_T$. Their one-loop $\beta$-function yielded a critical
point at $u^{*}_{m}= -1/\ln 2$ for $m>0$. (Upon including spin the
instabilities can occur in the charge or spin channels, the
critical coupling being $-1/2\ln 2$ for each channel.)  For $u_m >
-1/\ln 2$, the flow led to the Universal Hamiltonian, while for
$u_m<-1/\ln 2$ there was a runaway to strong coupling. The two
phases seemed clearly related to two types of behavior in level
spacings and the nature of quasiparticles. However, uncertainty
surrounding the nature of the strong-coupling phase and even its
very existence, predicated as it was on a one-loop calculation and
a fixed point of order unity, impeded further progress.

{\em We have verified  that the theory is  solvable in the limit
$g \to \infty$, with $1/g$ playing the role of $1/N$ in large-N
theories.}  The key idea is illustrated by proving that the four
point function $G_{\alpha \beta \gamma \delta}$ is just a sum of
repeated particle-hole bubble diagrams. Recall that in theories
with interaction $V=\lambda /N (\sum_i \psi^{\dag}_{i}\psi_i
)(\sum_j \psi^{\dag}_{j}\psi_j )$, only such iterated bubbles
survive, since only they have a free sum over $N$ for each extra
loop. The situation here is similar. Consider the one-loop
diagrams in our theory. Each vertex carries a sum over two momenta
( $\bk$ and $\bk'$ in Fig. 1a for the bare vertex), and each
propagator is diagonal in the disorder eigenvalue index, but not
the $\bk$ index.  In Fig. 1b,  the internal states  involve a sum
over terms of the form $ \phi^{*}_{\mu}(\bl )\phi^{}_{\mu}(\bm
)\phi^{*}_{\nu}(\bm )\phi^{}_{\nu}(\bk ) $ Replacing the sum (over
$\mu, \nu, \bl, \bm$ ) by the ensemble averages  using \beq
\langle \phi^{*}_{\mu}(\bk_1 )\phi^{}_{\mu}(\bk_2
)\phi^{*}_{\nu}(\bk_3 )\phi^{}_{\nu}(\bk_4) \rangle
={\delta_{12}\delta_{34}\over g^2} +O(1/g^3) \eeq (and neglecting
fluctuations down by $1/g$) we find that $\bm = \bl$ and that
there is a free sum over $g$ values of either one of them. (An
exception occurs when either vertex involves $u_0$ which does not
flow\cite{mm}.) The reader may  check that other possible one-loop
particle-hole  and particle-particle diagrams are down by $1/g$
because one or more of the external $\bk$ labels creep into the
diagram. The same logic holds for higher orders.
\begin{figure}
\narrowtext
\epsfxsize=2.4in\epsfysize=1.6in
\hskip 0.3in\epsfbox{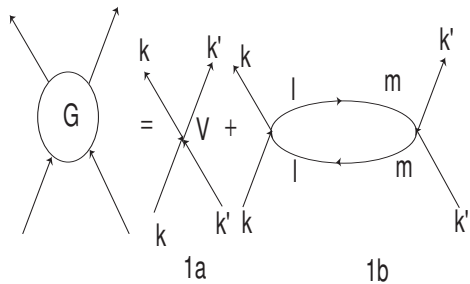}
\vskip 0.15in
\caption{Diagrammatic proof of the large-$N$ (with $N=g$)
nature of the theory.  Disorder-averaging internal lines forces
$\bl=\bm$ with a free sum.}
\label{large-n}
\end{figure}
Consequently, MM's one-loop $\beta$-function and critical exponent
$\nu= 1/\beta'(u^*) = 1$ are exact and the phase transition is
real. We now ask what it corresponds to.

 Let us  consider  just one
$u_m <0$ and  factorize: \begin{eqnarray*} \lefteqn{\exp
\left[{\Delta \over 2}\sum_{\bk \bk'}|u_m|\cos (m\theta -
m\theta')\right]=}\\
& &\int d\sigmab \exp \left[ -{|\sigmab|^2\over 2|u_m|
\Delta}-\bar{\psi}_{\alpha}\sigmab \cdot \bM_{\alpha
\beta}\psi_{\beta}\right]
\end{eqnarray*}
where $ \sigmab =(\sigma^1\ , \sigma^2 )$  has two components,  as
does $\bM$: \beq (M^1,M^2)=\sum_{\bk}\phi^{*}_{\alpha}(\bk
)\phi^{}_{\beta}(\bk )\left( \cos m \theta  \ , \sin m\theta
\right)\eeq

Integrating out the fermions we get an effective action \beq
S=-{|\sigmab|^2\over 2|u_m| \Delta} + Tr \ln \left[ (i\omega
-\varepsilon_\alpha )\bI -\sigmab \cdot \bM \right]
\label{effective}\eeq where $I$ is the unit matrix. The quadratic
part of the action is
\begin{eqnarray}S_0 &=&
 -\int d\tau \left[ {\dot{\sigmab}^2\over 4g\Delta^3}
+{|\sigmab|^2\over 2\Delta}\left[{1\over |u_m| }-{\ln 2
}\right]\right]\label{action}\end{eqnarray} upon using the
disorder-averaged result \beq \left< \sum_{\alpha
\beta}{N_\beta-N_\alpha\over
\varepsilon_\alpha-\varepsilon_\beta}\right> = {2 \over
\Delta^2}\int_{0}^{g\Delta\over2}
{d\varepsilon_1d\varepsilon_2\over
\varepsilon_1+\varepsilon_2}={2g \ln 2 \over \Delta}\eeq where
$N_{\alpha}$ is the Fermi occupation factor of level $\alpha$. The
$\dot{\sigmab}^2$ term is valid for frequencies well below
$\Delta$. (We shall not be  interested in faster  motion of the
collective mode.)

To reconfirm the large-$g$ nature of the theory,  one defines
$\bar{\sigmab}=\sigmab /g$ and evaluates the $Tr \ln$, term by
term, and finds that $S$ has a $g^2$ in front, which plays the
role of $1/\hbar$. If $ 1/g^2=''\hbar'' =0$, we are in the
classical limit and spontaneous symmetry breaking is possible
(even) for a single degree of freedom, while if $1/g^2 >0$, this
is impossible due to ``quantum'' fluctuations. If we include more
$u_m$'s, the corresponding $\sigmab$'s will make additive
contributions to $S$. As soon as one of them breaks symmetry, the
rest will not matter.

The phase diagram is shown in Figure 2. Let us first focus on
$1/g=0$.  For $u_m>u_{m}^{*}$, $\langle \sigmab \rangle =0$. For
$u_m<u_{m}^{*}$, $\langle \sigmab \rangle \ne 0$ with a Mexican
hat minimum located by balancing the quadratic term we have above
with the rest of $S$. This leads to $\langle |\sigmab| \rangle =
Cg\sqrt{|r|}$ where $C$ is a number of order unity and
$r=u_m^{-1}-(u_m^*)^{-1}$. At $1/g=0$ there are no fluctuations
and $\sigmab$ can sit anywhere on a circle.

What happens at  $1/g >0$ follows from its role as the pre-factor
in the action of $\bar{\sigmab}$:
 the critical point
can be felt  within a critical "fan" just as a $T=0$ critical
point can influence $T>0$ physics\cite{critical-fan}.  The fan is
$V$-shaped since both $r$ and $1/g$ scale with exponent $1$.

For $1/g>0$ the symmetry of the Mexican hat valley is {\em
explicitly} broken by sample-specific corrections to next order in
$1/g$ yielding a unique minimum at some angle $\theta$. The
stability of this minimum to   "quantum" fluctuations (at nonzero
$1/g$) is examined by reading off the hamiltonian for $\sigmab$
from Eqn.(\ref{action}) for the effective action. The angular part
of $\sigmab$  is governed by \beq H_\sigma ={L^2 \over
g}+gV(\theta ) \eeq where $L$ is the angular momentum conjugate to
$\theta$, and factors other than $g$ are suppressed. The radial
part of $<\sigmab>\simeq g$ which appears in both kinetic and
potential terms is represented by its average value of order $g$.
The  potential term dominates at large $g$ and ensures  that the
wavefunction is localized near the minimum of $V$. Unlike in the
two phases, the critical point (and fan region) has very large
fluctuations in $\sigmab$: From the hamiltonian \beq H_\sigma ={g
\bP^2 } + {\sigma^4\over g^2} \eeq we estimate the ground state
wave function to have a width ${\cal O} (g^{1/2})$ in $\sigma$. As
the interaction increases we cross over from the Universal
Hamiltonian phase through the critical region to the
symmetry-broken phase.

\begin{figure}
\narrowtext
\epsfxsize=2.4in\epsfysize=1.6in
\hskip 0.3in\epsfbox{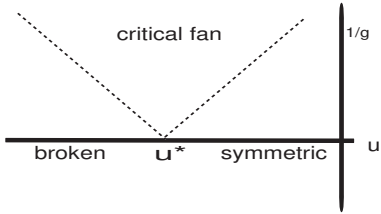}
\vskip 0.15in
\caption{ Schematic showing the critical point at $u_m^*=-1/\log{2}$ in the $g\to\infty$ limit on the $x$-axis.  As $g$ decreases one moves in the $y$-direction and the critical point controls the behavior of all physical quantities in a ``fan''.  }
\label{critical-fan}
\end{figure}

What does the order parameter correspond to? In a pure system when
any Landau parameter $u_m <-2$ ( $<-1$ with spin), the Fermi
surface undergoes a Pomeranchuk  shape
transition\cite{agd,oganesyan}. For example if $m=2$ the shape is
an oval\cite{oganesyan}, with the director being a Goldstone mode.
The present transition is the disordered version of this, with
small disorder terms making the Goldstone mode massive. (An
anisotropic gap at the Fermi surface could also arise when
considering instabilities in the spin channel\cite{varma}). Note
that the phenomenology described above can occur on top of a
mesoscopic Stoner transition\cite{andreev,brouwer,kurland}.

The  phase diagram clarifies  at least two issues. First, we can
relate the fluctuations of $\sigmab$ to ``delocalization in Fock
space''\cite{fock-loc}, which concerned the crossover of the
quasiparticle width to a Breit-Wigner form as a function of its
energy.  Note that  $ \sigmab$ acts as a background field for the
fermions in the Hartree -Fock Hamiltonian \beq H_{HF}
=\sum_{\alpha \beta } \psi^{\dag}_{\alpha}\left[
\varepsilon_{\alpha} \delta_{\alpha \beta} + \langle \sigmab
\rangle \cdot \bM_{\alpha \beta}\right] \psi_{\beta} .\label{hf}
\eeq Outside the critical fan $H_{HF}$ is controlled by one
sharply peaked value of $<\sigmab>$ (either zero or somewhere on
the Mexican hat's valley) and states are descendants of a single
Slater determinant. In the fan $\sigmab$ suffers large critical
fluctuations. Thus we expect the ground state to go from being
localized to delocalized and back to localized as we raise the
coupling at fixed $g$. This re-entrance has been seen in
numerics\cite{reentrant}.  We connect the quasiparticle
width\cite{fock-loc} to its energy $\varepsilon$, by arguing that
$\varepsilon$ plays the same role as $1/g$ or $T$, and that there
will be a fan in the $u-\varepsilon$ plane. As the quasiparticle
energy increases in either phase, one hits the crossover fan with
its delocalized states.

Second, we can understand the dependence of the CB peak spacing
distribution on $u$\cite{exact,hf}. $H_{HF}$ (Eqn.(\ref{hf}) is a
sum of two random Hamiltonians whose widths can be added in
quadrature to yield an effective level spacing \beq
\Delta'=\sqrt{\Delta^2+6<\sigmab>^2/g^2} \eeq In the weak-coupling
phase $<\sigmab>=0$ and there is no memory of $u_m$. In the broken
symmetric phase we have $<\sigmab>=C g\Delta\sqrt{r}$, and the
effective level spacing increases with $r$, and thus $u_m$. There
is a crossover between these two regimes in the critical fan.

Let us now briefly analyze the relevant experiments. The Sivan
{\it et al}\cite{sivan-expt} and Patel {\it et
al}\cite{marcus-expt} experiments are done on gated GaAs 2DEG
samples, for which $r_s\approx 1-1.2$ in the bulk. Using the area
of the sample and the fact that these dots are in the ballistic
limit we can find both $\Delta$ and $E_T$, and thence find
$g\approx 7-14$. Sivan {\it et al}\cite{sivan-expt} find that the
CB peak spacing is  about 5 times broader than that predicted by
the Universal Hamiltonian, which describes  the weak coupling
region of our phase diagram. However, Patel {\it et
al}\cite{marcus-expt} find it to be in accord with this
prediction, after accounting for ``experimental noise'' which is
determined by measuring the magnetic field asymmetry of the CB
spacings\cite{marcus-expt} (which indicates motion of the dopant
atoms, or some other scrambling of the single-particle potential).
Thus, the Patel {\it et al} data\cite{marcus-expt} seem to lie in
the weak-coupling region. Accounting for experimental noise could
also put the Sivan {\it et al} data in the weak-coupling region,
consistent with the equality of parameters in the two experiments.

The experiments by Simmel {\it et al} and Abusch-Magder {\it et
al}\cite{kastner-expt} are performed on Si quantum dots. Including
valley degeneracy one finds $r_s\approx2.2$, and $g\approx 18$.
While one cannot directly relate $r_s$ to $u_m$, one expects that
some $u_m$ might become more negative as $r_s$ increases. (For
$r_s\ge2$ local charge density correlations develop in the
dot\cite{wigner-like-th}, similar to the classical
limit\cite{shklovskii}. While a Fermi surface distortion is not a
charge density wave, it enhances the susceptibility for one, and
could thus be a precursor). Indeed, two signatures of the critical
fan are found in this experiment\cite{kastner-expt}. The CB
peak-spacing distribution is found to be 7-8 times wider than
expected from the Universal Hamiltonian (assuming spin
degeneracy). The most striking feature of the data is that the
width of the CB peaks does not vanish\cite{kastner-expt} as
$T\to0$. This is just what is expected for a system was located in
the critical fan: The ground and low-lying excited states are
``Fock-space delocalized'', and single-particle states are broad
even at low energy.

There are many problems one can attack using our approach. The
$d=2$ disordered, interacting bulk system, which shows
experimental\cite{2dmit} and theoretical\cite{belitz} indications
of undergoing a metal-insulator transition, with a Stoner-type
instability that occurs for arbitrary negative triplet
$u_{0t}$\cite{mucciolo}, can be modeled as a $d=2$ lattice of
quantum dots. In the strong-coupling phase, each dot will have a
slow collective degree of freedom, which could lead to
phase-breaking and lack of coherent backscattering of
quasiparticles at very low energies. Persistent current
measurements in systems with a flux, which have resisted
theoretical explanation\cite{persist}, may succumb to our
treatment, as may disordered gapless superconductors exhibiting a
novel metal-insulator transition\cite{sfbn}, since the eigenvalue
and eigenvector statistics of their grains are expected to be
governed by one of four newly discovered RMT universality
classes\cite{zirnbauer}. Effects of finite $T$ on the CB peak
spacing\cite{baranger} in the strong-coupling phase constitute a
natural extension of our work.

In summary, by identifying a small parameter $1/g$, we have managed to
control the problem of electrons subject to interactions and disorder
in ballistic quantum dots. At $1/g=0$ the (exact) saddle-point result
shows that order parameter $\sigmab$ acquires a nonzero average past a
critical coupling, and the correlation length exponent is $\nu=1$,
results coinciding with MM's results based on the one-loop
$\beta$-function\cite{mm}. The transition is the disordered version of
the Pomeranchuk transition of the Fermi
surface\cite{agd,oganesyan,varma}. For $1/g>0$ the critical point
controls a critical fan in which $\sigmab$ fluctuates wildly, leading
to a ground state not originating from a single Slater determinant,
and a broad quasiparticle peak even at low energy (which has been seen
in experiment\cite{kastner-expt}). Thus Fock-space delocalization is
just $\sigmab$ delocalization. The nonzero $\langle
\sigmab
\rangle \simeq {\cal O}(g) $ in the strong-coupling phase is seen
to explain the significant dependence of the CB peak spacing
distribution on the interaction, just as $\langle \sigmab \rangle =0 $
in the weak-coupling regime makes the distribution insensitive to
interaction. Finally, note that the increase of the effective level
spacing leads to an enhancement of the persistent
current\cite{persist1}.

We are grateful to the NSF for grants DMR-0071611 (GM), and DMR-
0103639 (RS), the Aspen Center for Physics for its hospitality,
and
 Sudip Chakravarty,
Claudio Chamon, Yong-Baek Kim, Harsh Mathur, Eduardo Mucciolo,
Chetan Nayak, Zvi Ovadyahu, Boris Shklovskii, and Chandra Varma
for illuminating conversations.


\begin{thebibliography}{99}
\bibitem{sivan-expt} U. Sivan {\it et al}, \prl\  {\bf 77}, 1123 (1996).
\bibitem{marcus-expt} S. R. Patel {\it et al}, \prl\ {\bf 80}, 4522 (1998).
\bibitem{kastner-expt} F. Simmel {\it et al}, \prb\ {\bf 59}, 10441 (1999);
D. Abusch-Magder {\it et al}, Physica E\ {\bf 6}, 382 (2000).
\bibitem{others-expt} F. Simmel, T. Heinzel, and D. A. Wharam,
Eur. Lett. \ {\bf 38}, 123 (1997); J. A. Folk {\it et al},
\prl\ {\bf 86}, 2102 (2001); S. L\"uscher {\it et al}, \prl\ {\bf 86}, 2118 (2001).
\bibitem{alt1} K. B. Efetov, Adv. Phys. {\bf 32}, 53 (1983);
B. L. Al'tshuler ad B. I. Shklovskii, { Sov. Phys. JETP}\
{\bf 64}, 127 (1986).
\bibitem{RMT} M. L. Mehta, {\it Random Matrices}, Academic Press, San
Diego, 1991.
\bibitem{reviews} For recent reviews, see, T. Guhr, A. M\"uller-Groeling,
and H. A. Weidenm\"uller, Phys. Rep. {\bf 299}, 189 (1998);
Y. Alhassid, \rmp\ {\bf 72}, 895 (2000); A. D. Mirlin, Phys. Rep. {\bf
326}, 259 (2000).
\bibitem{2brim} J. B. French and S. S. M. Wong, Phys. Lett. {\bf 33B},
447 (1970); O. Bohigas and J. Flores, Phys. Lett. {\bf 34B}, 261
(1971); Y. Alhassid, Ph. Jacquod, and A. Wobst, \prb\ {\bf 61}, 13357
(2000); Physica E {\bf 9}, 393 (2001); Y. Alhassid and A. Wobst,
\prb\ {\bf 65}, 041304 (2002).
\bibitem{brouwer} P. W. Brouwer, Y. Oreg, and B. I. Halperin, \prb\
{\bf 60}, 13977 (1999).
\bibitem{kurland} I. L. Kurland, I. L. Aleiner, and B. L. Al'tshuler,
\prb\ {\bf 62}, 14886 (2000).
\bibitem{univ-ham}I. L. Aleiner, P. W. Brouwer, and L. I. Glazman,
Phys. Rep. {\bf 358}, 309 (2002), and references therein; Y. Oreg, P. W. Brouwer,
X. Waintal, and B. I. Halperin, cond-mat/0109541, and references
therein.
\bibitem{mm} G. Murthy and H. Mathur, \prl\ {\bf 89}, 126804 (2002).
\bibitem{rg-shankar} R. Shankar, {\it Physica}\ {\bf A177},
530 (1991); R.Shankar, { Rev. Mod. Phys.} {\bf 66}, 129 (1994).
\bibitem{agd} A. A.  Abrikosov, L. P. Gorkov, and I. E. Dzyaloshinski,
{\it Methods of Quantum Field Theory in Statistical Physics}, Dover
Publications, New York, 1963.
\bibitem{rg-us} T. Tokuyasu, M. Kamal, and G. Murthy,
\prl\ {\bf 71}, 4202 (1993); N. Berdenis and G. Murthy,
\prb\ {\bf 52}, 3083 (1995); G. Murthy and S. Kais,
{ Chem. Phys. Lett.}\ {\bf 290}, 199 (1998).
\bibitem{critical-fan} S. Chakravarty, B. I. Halperin, and
D. R. Nelson, \prl\ {\bf 60}, 1057 (1988); \prb\ {\bf 39}, 2344
(1989); For a detailed treatment of the generality of the phenomenon,
see, S. Sachdev, {\it Quantum Phase Transitions}, Cambridge University
Press, Cambridge 1999.
\bibitem{oganesyan} V. Oganesyan, S. A. Kivelson, and E. Fradkin, \prb\ {\bf 64}, 195109 (2001).
\bibitem{varma} C. M. Varma, \prl \ {\bf 83}, 3538 (1999).
\bibitem{andreev} A. V. Andreev and A. Kamenev, \prl\ {\bf 81}, 3199 (1998).
\bibitem{fock-loc} B. L. Al'tshuler, Y. Gefen, A. Kamanev,
and L. S. Levitov, \prl\ {\bf 78}, 2803 (1997).
\bibitem{reentrant} R. Berkovits and Y. Avishai, \prl \ {\bf 80}, 568 (1998).
\bibitem{exact} R. Berkovits, \prl\ {\bf 81}, 2128 (1998).
\bibitem{hf} A. Cohen, K. Richter, and R. Berkovits, \prb\ {\bf 60}, 2536 (1999);
P. N. Walker, G. Montambaux, and Y. Gefen, {\it ibid}, 2541 (1999);
S. Levit and D. Orgad, \prb\ {\bf 60}, 5549 (1999); D. Ullmo and
H. U. Baranger, \prb\ {\bf 64}, 245324 (2001); V. Belinicher,
E. Ginossar, and S. Levit, cond-mat/0109005; Y. Alhassid and
S. Malhotra, cond-mat/0202453.
\bibitem{wigner-like-th} P. N. Walker, Y. Gefen, and G. Montambaux, \prl\
{\bf 82}, 5329 (1999); K.-H. Ahn, K. Richter, and I.-H. Lee, \prl\
{\bf 83}, 4144 (1999).
\bibitem{shklovskii} A. A. Koulakov, F. G. Pikus, and
B. I. Shklovskii, \prb\ {\bf 55}, 9223 (1997); A. A. Koulakov and
B. I. Shklovskii,
\prb\ {\bf 57}, 2352 (1998); Phil. Mag.  {\bf B77},  1235 (1998).
\bibitem{2dmit}V. M. Pudalov,  M. D'Iorio, S. V. Kravchenko,
and J. W. Campbell, \prl\ {\bf 70}, 1866 (1993).
\bibitem{belitz} For a review of the theory, see, D. Belitz and
T. R. Kirkpatrick, { Rev. Mod. Phys.} \ {\bf 66}, 261 (1994).
\bibitem{mucciolo} C. de C. Chamon and E. Mucciolo, \prl {\bf 85}, 5607 (2000).
\bibitem{persist} U. Eckern and P. Schwab, J. Low Temp. Phys. {\bf 126}, 1291 (2002).
\bibitem{sfbn} T. Senthil, M. P. A. Fisher, L. Balents, and C. Nayak, \prl {\bf 81}, 4704 (1998).
\bibitem{zirnbauer} M. R. Zirnbauer, J. Math. Phys. {\bf 37}, 4986 (1996).
\bibitem{baranger} G. Usaj, and H. U. Baranger, \prb\ {\bf 64}, 201319 (2001); cond-mat/0203074.
\bibitem{persist1} B. L. Al'tshuler, Y. Gefen, and Y. Imry,
\prl {\bf 66}, 88 (1991).
\end{thebibliography}
 \end{document}